\documentclass{rspublic}

\usepackage{graphicx}

\begin{document}

\title[RG and nuclear forces]{The renormalisation group and\\ nuclear forces}

\author[M. C. Birse]{Michael C. Birse}

\affiliation{School of Physics and Astronomy, The University of Manchester,
Manchester, M13 9PL, U.K.}

\label{firstpage}

\maketitle

\begin{abstract}{renormalisation group, nuclear forces, effective field theory}
I give an outline of recent applications of the renormalisation group to effective 
theories of nuclear forces, focussing on the use of a Wilsonian approach to analyse
systems of two or three nonrelativistic particles.
\end{abstract}

\section{Introduction}

The last twenty years have seen a major shift in how we describe the forces between 
nucleons. This was triggered by the suggestion of Weinberg (1990, 1991) that the ideas 
of effective field theory (EFT) could be applied to these strongly interacting 
systems. This approach offers the possibility of a systematic, model-independent
treatment which, ultimately, could form a bridge to the underlying theory of the 
strong interaction, Quantum Chromodynamics (QCD).

Effective field theories are built out of fields corresponding to the appropriate
low-energy degrees of freedom and contain all possible terms consistent with the
symmetries of the underlying dynamics. To have predictive power, such a theory 
must be expandable in powers of ratios of low-energy scales to those of the
underlying physics. For nuclear physics, the low-energy scales, denoted generically 
by $Q$, include the momenta of the nucleons and the mass of the pion. The 
underlying scales of QCD, denoted by $\Lambda_0$, include $4\pi F_\pi$ (the scale
associated with the hidden chiral symmetry) as well as the masses of the nucleons 
and $\rho$, $\omega$ and other mesons.

Since Weinberg made his original proposal, debate has raged within the community 
over the appropriate power counting to use in organising this expansion in 
powers of low-energy scales. Extensive reviews can be found in the articles by 
Beane \textit{ et al.}~(2001), Bedaque \& van Kolck (2002) and Epelbaum 
\textit{ et al.}~(2009), and more recent summaries from two different points of 
view have been given by Epelbaum and Gegelia (2009) and Birse (2009). The main 
bones of contention have been: how do we renormalise nonperturbative systems 
consistently, and which pieces of the interaction should we iterate to all orders?

To get unambiguous answers to these questions we need a rigorous tool to
analyse the scale dependences of physical systems, and this is provided by
the renormalisation group (RG). I concentrate here on the version developed in 
Manchester (Birse \textit{et al.}~1999; Barford \& Birse, 2003, 2005; 
Birse, 2006a, 2006b) which uses a Wilsonian approach to construct functional RG 
equations. These can be solved exactly, at least in simple cases. However many 
of the results discussed here were first obtained from more heuristic 
RG equations with momentum-space cutoffs (Lepage 1997;
Bedaque \& van Kolck 1998; van Kolck 1999; Bedaque \textit{et al.}~1999a, 
1999b, 2000, 2003a; Platter \textit{et al.}~2004; 
Nogga \textit{et al.}~2005; Griesshammer 2005; Platter \& Hammer 2007), 
by using dimensional regularisation with subtraction of power-law divergences 
(Kaplan \textit{et al.}~1998a, 1998b; Phillips \textit{et al.}~1999;
Kong \& Ravndal 1999, 2000; Ando \& Birse 2008), or with a 
Bogoliubov-Parasiuk-Hepp-Zimmermann subtractive renormalisation scheme 
(Gegelia, 1999a, 1999b).

Other closely related approaches that are being explored include extensions
of the subtractive renormalisation scheme (Frederico \textit{et al.}~1999, 2000; 
Hammer \& Mehen 2001; Afnan \& Phillips 2004; Tim\'oteo \textit{et al.}~2005, 2010; 
Yang \textit{et al.}~2008, 2009a, 2009b) and the use of a radial cutoff in 
coordinate space (Pav\'on Valderrama \& Ruiz Arriola 2004a, 2004b, 2006a, 2006b, 
2008, 2009). 

In addition, there is a powerful functional RG based on the 
Legendre-transfor\-med effective action (Wetterich 1993; Berges \textit{et al.}~2002).
This has recently been applied to few-body (as well as many-body) systems of 
strongly interacting, nonrelativistic particles (Birse \textit{et al.}~2005, 2010a,
2010b; Diehl \textit{et al.}~2007, 2008; Birse 2008; Floerchinger 
\textit{et al.}~2009; Moroz \textit{et al.}~2009; Schmidt \& Moroz 2010). 
As generally implemented, this is less rigorous than the versions of the RG 
mentioned above because it relies on truncations of the effective action that are 
not based on any systematic power counting. None-the-less it is proving very useful 
in situations where exact solutions of other RG equations cannot be found, such as 
four-body systems and dense matter. 

Lastly, I should mention the approach known as $V_{\mbox{\scriptsize low-}k}$ 
(Bogner \textit{et al.}~2003a, 2003b) and related applications of the similarity 
renormalisation group to nuclear forces (Bogner \textit{et al.}~2007a, 2007b;
Jurgenson \textit{et al.}~2008, 2009). Both of these lead to effective 
interactions that evolve according to RG equations. However the structures 
of the resulting equations are more complicated than those of the 
approaches mentioned above and, at least so far, it has not been possible to 
analyse their scaling behaviours.

Whatever regulator or subtraction scheme we choose, we should first identify 
fixed-point solutions of the RG equations. If we expand a general solution around 
one of these points, we can use the linearised RG equations to classify the
perturbations as relevant, marginal or irrelevant. The eigenvalues of the 
linear equations give the anomalous dimensions of the corresponding operators 
and hence can be used to construct a power-counting scheme. This counting is 
what makes it possible to define a systematic expansion of the corresponding EFT.

In systems whose particles interact very strongly at low energies, the terms in 
the potential can have large anomalous dimensions. Their scaling behaviour, 
and hence the power counting for the corresponding EFT, is then quite different 
from naive dimensional analysis. This is particularly true of systems close to 
the ``unitary limit", that is, with very large two-body scattering lengths. Nuclear
forces are a prime example of this, since the nucleon-nucleon scattering lengths
are of the order of 5--20~fm, much larger than the range of the interactions.

In such cases, the leading two-body interaction can become a relevant term, and 
other two-body operators are strongly promoted compared to naive expectations.
The two-body sectors of the resulting effective theories are really just versions 
of the much older effective-range expansions (Bethe 1949; 
Blatt \& Jackson 1949; van Haeringen \& Kok 1982; Badalyan \textit{et al.}~1982). 
The benefit of the modern field-theoretic framework is that it can provide 
consistent effective current operators as well as three- and more-body forces. 

The first application of RG methods to three-body forces in theories with 
two-body contact interactions was to channels of three fermions with 
mixed-symmetry spatial wave functions. These give rise to a repulsive 
``particle-exchange" force between an interacting pair and the third particle 
(Bedaque \& van Kolck, 1998; Bedaque, Hammer \& van Kolck, 1998). This work 
showed that three-body forces were not required to describe the low-energy physics 
in these systems, implying that there are no large, negative anomalous dimensions.
Subsequent, more detailed RG analyses in momentum space (Griesshammer 2005)
and in coordinate space (Birse 2006b) have determined the exact anomalous 
dimensions for these repulsive three-body channels. 

In systems of three bosons or three fermions with a fully symmetric spatial 
wave function, the particle-exchange force is attractive. This can have 
dramatic consequences, most notably the Efimov effect (Efimov 1971, 1979):
a tower of bound states with energies in a geometric sequence. The RG
flow in these systems displays a limit-cycle behaviour (Bedaque, Hammer \& 
van Kolck, 1999a, 1999b, 2000; G\l azek and Wilson, 2004; Barford \& Birse, 2005;
Mohr \textit{et al.}, 2006),
which corresponds to anomalous breaking of scale invariance to a discrete remnant
of the symmetry. The leading three-body force in these systems is a marginal
term, corresponding to the starting point on the limit cycle. This means that 
one piece of three-body physics is sufficient to determine their low-energy
behaviour, as has long been known from the Phillips line (Phillips 1968, 1977)
which shows a correlation between the triton binding energy and $nd$ scattering 
length for forces fitted to the two-body scattering data.

In contrast, four-body systems show no evidence for relevant or marginal 
forces, even in cases where the three-body subsystems display Efimov 
behaviour (Platter, Hammer \& Meissner, 2004, 2005; Platter \& Hammer, 2007).
This means that their low-energy observables are determined purely by
two- and three-body physics, providing an explanation for the ``Tjon line" 
correlation between $^4$He and triton binding energies (Tjon 1975).
No exact results are available for the scaling behaviour in four-body systems,
but functional RG methods are now being used to estimate anomalous dimensions 
for both bosonic and fermionic systems (Schmidt \& Moroz, 2010; Birse 
\textit{et al.}, 2010b).

All of the RG techniques mentioned above are applicable to any system of 
strongly interacting, nonrelativistic particles, where they can be used to 
determine the pertinent power counting and hence to set up a consistent EFT 
description. Other areas where these ideas are now being applied within 
nuclear and hadron physics are weakly bound ``halo" nuclei (Bertulani 
\textit{et al.}~2002; Bedaque \textit{et al.}~2003b; Higa 
\textit{et al.}~2008) and the interactions 
of mesons containing heavy quarks (Braaten \& Kusonoki 2004; Fleming 
\textit{et al.}~2007; Braaten \textit{et al.}~2010; Hagen \textit{et al.}~2010).
Looking further afield, they are being used increasingly in studies of 
ultracold atomic systems (see, for example: Braaten \& Hammer 2006).

\section{Scales and scaling}

Underpinning any viable EFT is an expansion in powers of the low-energy
scales for a system. The RG can help to elucidate this scale dependence 
but, first, we need to identify all the relevant scales, $Q$. 
The most obvious ones for any low-energy theory are particles' momenta,
both on-shell and off-shell. 

Typical momenta in nuclear systems are often of the order of 100 or 200~MeV,
which is comparable to the mass of the pions. These mesons are not only the 
approximate Goldstone bosons of QCD, they also give rise to the longest-range 
forces between nucleons. If we wish to describe physics on this scale, we need 
to include pions in our EFT, and to count their mass as one of our low-energy 
scales. 

So far, this list of scales is just the same as in chiral perturbation theory 
(ChPT), Weinberg's original EFT for low-energy meson physics (Weinberg, 1979).
However, in contrast to the case of pions, where the hidden chiral symmetry ensures 
that their interactions are weak at low energies, nucleons interact strongly, 
forming bound states (nuclei). 

\begin{figure}[h]
\begin{center}
\includegraphics[width=2cm]{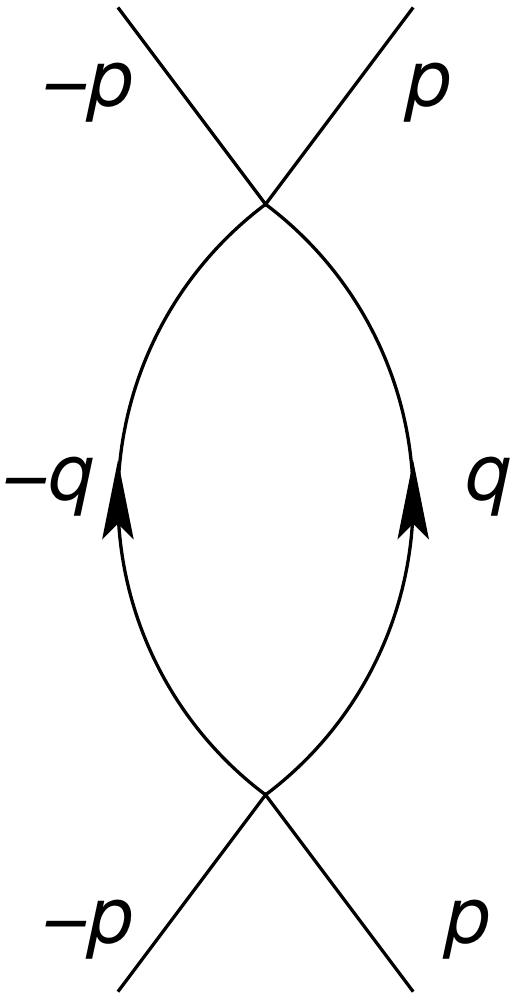}
\end{center}
\caption{Loop diagram for two-body scattering}
\end{figure}

To see why this this can lead to problems with extending ChPT to two 
or more nucleons, we should look at the nonrelativistic loop diagram for 
NN scattering in figure 1. For contact interactions, this has the form
\begin{equation}
\frac{M}{(2\pi)^3}\int \frac{{\rm d}^3q}{p^2-q^2+{\rm i}\epsilon}
=-{\rm i}\,\frac{M\,p}{4\pi}+\mbox{analytic in}\ p^2.
\label{eq:basicloop}
\end{equation}
As noted by Weinberg (1990, 1991), this
is enhanced to order $Q$, instead of $Q^2$ as in the relativistic case. 
Nonetheless, the leading terms of the potential are of order $Q^0$ (OPE 
and the simplest contact interaction) and so each iteration 
is suppressed by a power of $Q/\Lambda_0$. The theory is therefore still
perturbative, provided $Q<\Lambda_0$.

In fact the analytic part of the integral (\ref{eq:basicloop}) is linearly 
divergent and so we need to either cut it off or subtract it at some scale 
$q=\Lambda$. Iterating the potential then leads to contributions with
powers of $\Lambda/\Lambda_0$. These will again be perturbative provided we
keep our cutoff within the domain of our EFT, 
$\Lambda<\Lambda_0$, and so they cannot generate bound states.

We therefore need to identify further low-energy scales. Of particular 
interest are any that promote some of the interactions to order $Q^{-1}$ 
(making them marginal terms in RG language) since these can, and indeed must, be 
treated nonperturbatively. The first examples to be identified in NN scattering 
were provided by the $S$-wave scattering lengths (Bedaque \& van Kolck 1998; 
van Kolck 1999; Kaplan \textit{et al.}~1998a, 1998b; Birse \textit{et al.}~1999).

In addition, there are scales associated with long-range interactions. A simple 
example is provided by the Coulomb potential between two charged particles, such 
as two protons. Here, after we scale the nucleon mass $M_{\scriptscriptstyle N}$
out of the Hamiltonian, the strength of interaction can be expressed in terms of 
the inverse Bohr radius,
\begin{equation}
\kappa=\frac{\alpha\,M_{\scriptscriptstyle N}}{2}\simeq 3.4\;\mbox{MeV}.
\end{equation}

The long-range pion-exchange forces can be expanded using the methods of ChPT. 
The leading piece is one-pion exchange (OPE) whose strength can be expressed in 
terms of the momentum scale
\begin{equation}
\lambda_{pi{\scriptscriptstyle NN}}
=\frac{16\pi F_\pi^2}{g_{\scriptscriptstyle A}^2 
M_{\scriptscriptstyle N}}\simeq 290\;\mbox{MeV}.
\end{equation}
This is built out of high-energy scales in chiral perturbation theory, 
$4\pi F_\pi$ and $M_{\scriptscriptstyle N}$. Counting it as a high-energy 
scale, would lead us to a perturbative treatment of OPE, as developed by 
Kaplan \textit{et al.}~(1998a, 1998b). Numerically, however, the value of 
$\lambda_{\pi{\scriptscriptstyle NN}}$ is only about twice $m_\pi$, 
suggesting that it may be better viewed as a low-energy scale. If we do
so, OPE is promoted to order $Q^{-1}$, implying that it
should be iterated. 

Following a Wilsonian approach, the next step in the RG is to cut off our
EFT at some arbitrary scale $\Lambda$, lying
above the low-energy scales $Q$ but below the scale $\Lambda_0$
of the underlying physics, as in figure~2. (This assumes good separation 
of these scales, as required for an EFT with a convergent expansion.)
\begin{figure}[h]
\begin{center}
\includegraphics[width=5cm]{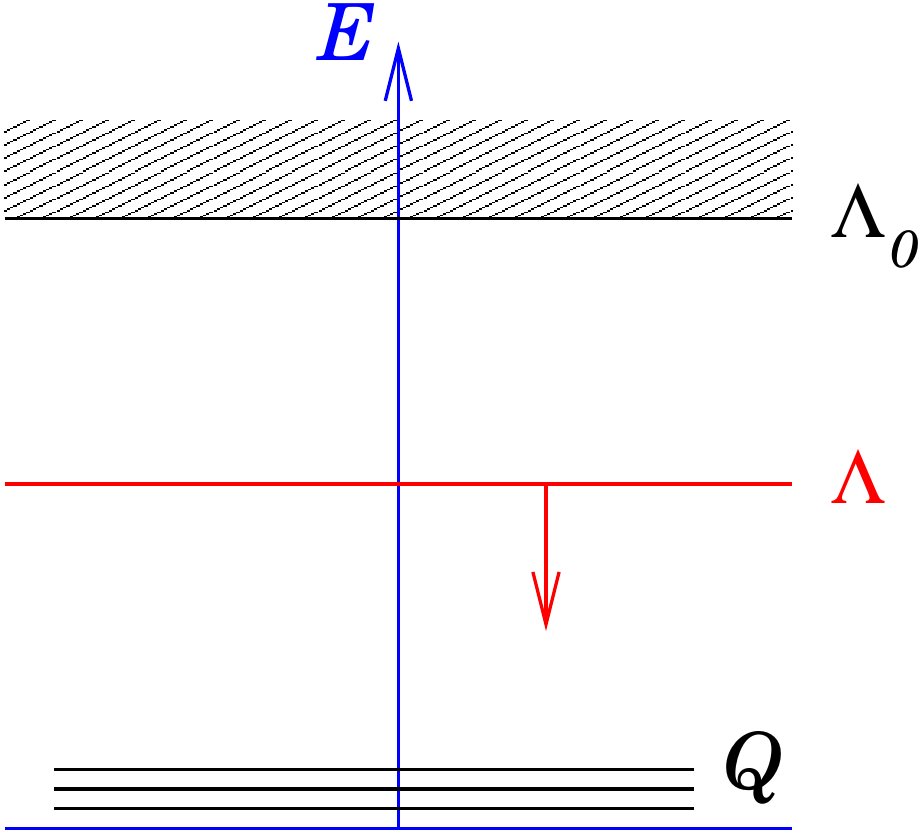}
\end{center}
\caption{The running cutoff $\Lambda$.}
\end{figure}

Then we can follow the evolution of our theory as we ``integrate out'' more 
and more of the physics by lowering $\Lambda$. As we vary our arbitrary cutoff,
we demand that physics (for example, the scattering matrix) be independent of 
$\Lambda$. This means that the couplings in our EFT must run with $\Lambda$ 
to compensate for the physics we are integrating out. Ultimately, for 
$\Lambda\ll\Lambda_0$, we lose all memory of the underlying physics 
and the only scale left is $\Lambda$. 

\begin{figure}[h]
\centering
\includegraphics[width=6cm]{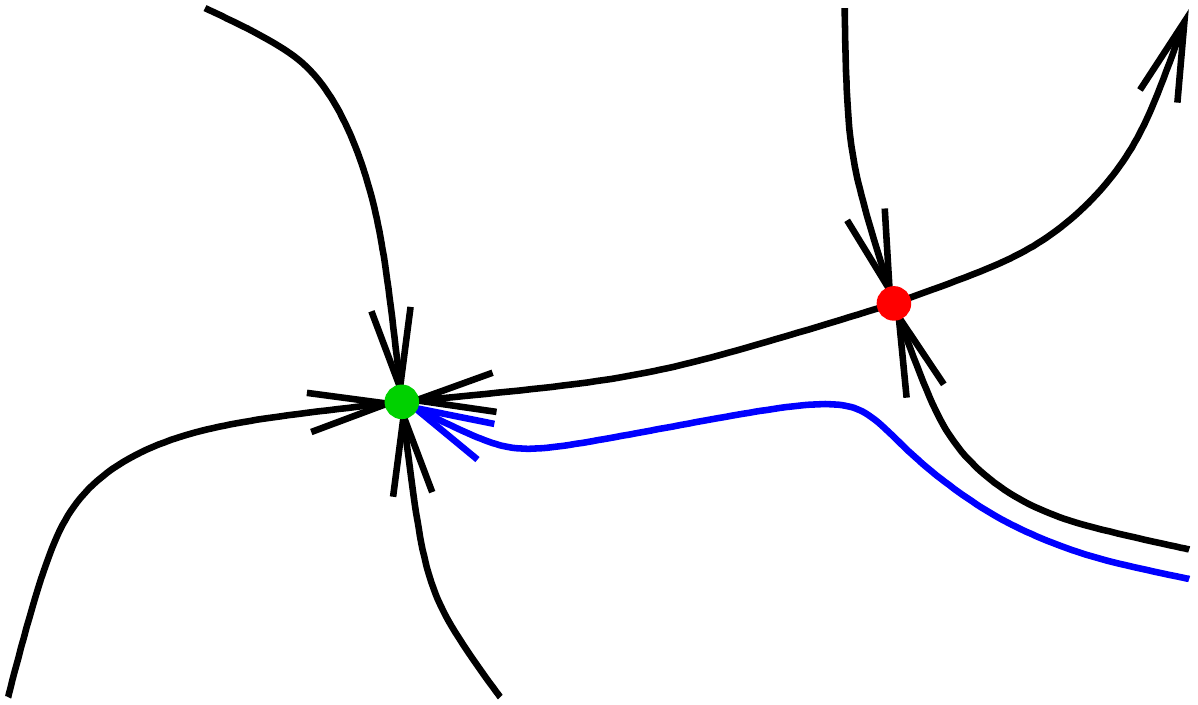}
\caption{Fixed points of an RG flow.}
\end{figure}

Finally, we rescale the theory by expressing all dimensioned quantities in 
units of $\Lambda$. At one of the end points of the RG flow, all couplings 
are then just numbers independent of $\Lambda$. We have arrived at a fixed 
point of the RG: a theory that describes a scale-free system. Two are 
shown in figure 3. The one on the left is stable: any nearby theory will 
flow towards it as the the cut-off is lowered. In contrast, the one on the 
right has an unstable direction: the flow can take theories away from the 
fixed point unless they lie on the ``critical 
surface".

Close to a fixed point, we can find perturbations that scale with definite 
powers of $\Lambda$. They can be classified into three types:
\begin{itemize}
\item $\Lambda^{-\nu}$: relevant (super-renormalisable in the language of
particle physics),\newline
for example mass terms in quantum field theories like QED;
\item $\Lambda^0$: marginal (renormalisable),\newline
for example the couplings familiar in gauge theories like the Standard Model
(typically these show a $\log\Lambda$ dependence on the cut-off);
\item $\Lambda^{+\nu}$: irrelevant (nonrenormalisable),\newline
for example the interactions in mesonic ChPT.
\end{itemize}
We now can expand our EFT around one of the fixed points using these perturbations.
Their RG scaling maps directly onto the order in the usual power counting (Weinberg, 1979), an interaction running as $\Lambda^\nu$ corresponding to a term in the EFT 
of order $Q^d$ where $d=\nu-1$. 

\section{Short-range forces}

To illustrate how RG methods can be applied to scattering of nonrelativistic 
particles, let us look at a system of two particles at energies where the 
range of the forces is not resolved (for example, two nucleons with an energy 
below about 10~MeV). This can be described by an effective 
Lagrangian with two-body contact interactions or, equivalently, a Hamiltonian
with $\delta$-function potentials. In momentum space, the $S$-wave potential can 
be written
\begin{equation}
V(k',k,p)=b_{00}+b_{20}(k^2+k'^2)+b_{02}\,p^2\cdots,
\end{equation}
where $k$ and $k'$ denote the initial and final relative momenta and the 
energy-dependence is expressed in terms of the on-shell
momentum $p=\sqrt{ME}$. 

Scattering can be described by the reactance matrix, defined similarly to 
the scattering matrix but with standing-wave boundary conditions. This has 
the advantage that it is real below the particle-production threshold. For 
$S$-wave scattering, it satisfies the Lippmann-Schwinger equation
\begin{equation} 
K(k',k,p)=V(k',k,p)
+\frac{M}{2\pi^2}\,{\cal{P}}\!
\int_{0}^{\Lambda}q^2{\rm d}q\,\frac{V(k',q,p)K(q,k,p)}
{{p^2}-{q^2}},
\end{equation}
where ${\cal P}$ denotes the principal value. This integral equation sums 
chains of the bubble diagrams in figure 1 to all orders. 

With contact interactions, the integral over the momentum $q$ of the 
virtual states is divergent and so we need to regulate it. Here I follow the 
method developed by Birse \textit{et al.}~(1999) and simply cut the
integral off at $q=\Lambda$. 
Demanding that the off-shell $K$-matrix be independent of $\Lambda$,
\begin{equation}
\dot K\equiv \frac{\partial K}{\partial\Lambda}=0,
\end{equation}
ensures that scattering observables will be independent of the
arbitrary cut-off. If we write the integral equation for $K$ in the 
schematic form
\begin{equation}
K=V+VGK,
\end{equation}
differentiating it gives
\begin{equation}
0=\dot V+\dot VGK+V\dot GK,
\end{equation}
where $\dot G$ implies differentiation with respect to the cut-off on the
integral. Since this involves the off-shell $K$ matrix, we can use the
integral equation for $K$ to convert it into the form
\begin{equation}
\dot V=-V\dot GV.
\end{equation}
This describes the evolution of the potential as the cutoff is lowered
and states are integrated out of the effective theory. Written out explicitly, 
it is
\begin{equation}
\frac{\partial V}{\partial\Lambda} 
=\frac{M}{2\pi^2}\,V(k',\Lambda,p,\Lambda)\,\frac{\Lambda^2}{\Lambda^2-p^2}\,
V(\Lambda,k,p,\Lambda).
\end{equation}
Note that the use of the fully off-shell $K$ matrix was essential to
obtaining an equation involving only the potential. The corresponding
equation for the evolution of $V_{\mbox{\scriptsize low-}k}$ 
(Bogner \textit{et al.}~2003a, 2003b) is based on the half-off-shell 
$T$ matrix and so still involves the scattering matrix.

Our equation for the cutoff dependence of the effective potential is 
still not quite an RG equation: the final step is to express all dimensioned
quantities in units of $\Lambda$. Rescaled momentum variables (denoted with hats)
are defined by $\hat k=k/\Lambda$ etc., and a rescaled potential by
\begin{equation}
\hat V(\hat k',\hat k,\hat p,\Lambda)=\frac{M\Lambda}{2\pi^2}\,
V(\Lambda\hat k',\Lambda\hat k,\Lambda\hat p,\Lambda).
\end{equation}
(The factor $M$ in this corresponds to dividing an overall factor of $1/M$ out 
of the Schr\"odinger equation.) This satisfies the RG equation 
\begin{eqnarray}
\Lambda\,\frac{\partial\hat V}{\partial\Lambda}&=&
\hat k'\,\frac{\partial\hat V}{\partial\hat k'}
+\hat k\,\frac{\partial\hat V}{\partial\hat k}
+\hat p\,\frac{\partial\hat V}{\partial\hat p}
+\hat V\cr
&&\qquad+\hat V(\hat k',1,\hat p,\Lambda)\,\frac{1}{1-\hat p^2}\,
\hat V(1,\hat k,\hat p,\Lambda).
\label{rgeqn}
\end{eqnarray}
The sum of logarithmic derivatives is similar to the structure of analogous
RG equations used in other areas of physics; it counts the powers of
low-energy scales present in the potential. The boundary conditions on 
its solutions are that they should be analytic functions of 
$\hat k^2$, $\hat k^{\prime 2}$ and $\hat p^2$ (since they should arise from
an effective Lagrangian constructed out of $\partial/\partial t$ and 
$\nabla^2$). 

Having constructed our RG equation, the first thing we need to do is to
look for its fixed points -- solutions that are independent of $\Lambda$.
Let us start with the obvious one, the trivial fixed point:
\begin{equation}
\hat V=0.
\end{equation}
Since this gives no scattering, it obviously describes a scale-free system.

To describe more interesting physics, we need to expand around the fixed point,
looking for perturbations that scale with definite powers of $\Lambda$. These
are eigenfunctions of the linearised RG equation. They have the form
\begin{equation}
\hat V(\hat k',\hat k,\hat p,\Lambda)=\Lambda^\nu \phi(\hat k',\hat
k,\hat p),
\end{equation}
and they satisfy the eigenvalue equation
\begin{equation}
\hat k'\,\frac{\partial\phi}{\partial\hat k'}
+\hat k\,\frac{\partial\phi}{\partial\hat k}
+\hat p\,\frac{\partial\phi}{\partial\hat p}
+\phi=\nu\phi.
\end{equation}
Its solutions are
\begin{equation}
\phi(\hat k',\hat k,\hat p)=C\,\hat k^{\prime 2l}\,\hat k^{2m}\,\hat p^{2n},
\end{equation}
with $k,l,m\geq 0$ since only non-negative, even powers satisfy the boundary
condition. The corresponding eigenvalues are
\begin{equation}
\nu=2(l+m+n)+1.
\end{equation}
These are all positive and so the fixed point is stable. The eigenvalues
$\nu$ simply count the powers of low-energy scales and can be written 
$\nu=d+1$ where $d$ is the ``engineering dimension" of an operator in 
the potential.

In addition, there are various nontrivial fixed points, all of which are unstable. 
The most interesting one is purely energy-dependent. To study it, let us focus
on potentials of the form $V(p,\Lambda)$. The RG equation for these 
can be rewritten in the form of linear equation for $1/\hat V(\hat p,\Lambda)$,
\begin{equation}
\Lambda\,\frac{\partial}{\partial\Lambda}\left(\frac{1}{\hat V}\right)
=\hat p\,\frac{\partial}{\partial\hat p}\left(\frac{1}{\hat V}\right)
-\frac{1}{\hat V}-\frac{1}{1-\hat p^2}.
\end{equation}

To find its fixed point, we set the LHS of this equation to zero. The
resulting ODE can then be integrated easily. The solution that satisfies
our boundary condition is
\begin{equation}
\frac{1}{\hat V_0(\hat p)}=-1+\frac{\hat p}{2}\ln\frac{1+\hat p}{1-\hat p}.
\end{equation}
The precise form of this is regulator-dependent, but the presence
of a negative constant of order unity is universal.

Since this potential has no momentum dependence, the integral equation for 
$K$ simplifies to an algebraic equation. In rescaled, dimensionless
form, it can be written
\begin{equation}
\frac{1}{\hat K(\hat p)}=\frac{1}{\hat V_0(\hat p)}-\int_0^1
\frac{\hat q^2\,{\rm d}\hat q}{\hat p^2-\hat q^2}.
\end{equation}
The integral here is, up to a sign, the same as the one in $1/\hat V_0$ itself 
and hence we get
\begin{equation}
\frac{1}{\hat K(\hat p)}=0.
\end{equation}
The corresponding $T$ matrix has a pole at $p=0$ and so the fixed-point 
describes a system with a bound state at exactly zero energy. This is often
called the ``unitary limit" of two-body scattering and it forms another
example of a scale-free system.

More general systems can be described by perturbing around the fixed point. 
In particular, energy-dependent perturbations can be found by substituting 
\begin{equation}
\frac{1}{\hat V(\hat p,\Lambda)}=\frac{1}{\hat V_0(\hat p)}
+\Lambda^\nu\phi(\hat p)
\end{equation}
into the RG equation. The functions $\phi(\hat p)$ satisfy the eigenvalue 
equation
\begin{equation}
\hat p\,\frac{\partial\phi}{\partial\hat p}-\phi=\nu\phi.
\end{equation}
The solutions to this are powers of the energy,
\begin{equation}
\phi(\hat p)=C\hat p^{2n},
\end{equation}
with eigenvalues
\begin{equation}
\nu=2n-1.
\end{equation}
The RG eigenvalues for these perturbations have been shifted by $-2$ compared
to the simple ``engineering" power counting. There is one negative eigenvalue
and so the fixed point is unstable. 

A slice through the RG flow for equation (\ref{rgeqn}) is shown in figure 4. 
The two fixed points can be seen, as well as the critical line through the 
nontrivial one. As in the generic flow in figure 3, potentials close to this 
line initially flow towards the fixed point as we lower the cut-off but are then 
diverted away from it. A potential to the right of the line is not quite strong 
enough to produce a bound state. As $\Lambda$ passes through the scale associated 
with the virtual state, the flow turns to approach the trivial fixed point from the 
weakly attractive side. In contrast, a potential to the left of the critical line
generates a finite-energy bound state. This state drops out of our low-energy 
effective theory when the cut-off reaches the corresponding momentum scale. As 
this happens, the RG flow takes the potential to infinity and it then reappears 
from the right, ultimately approaching the trivial fixed point from the weakly 
repulsive side.

\begin{figure}[h]
\centering
\includegraphics[width=9cm]{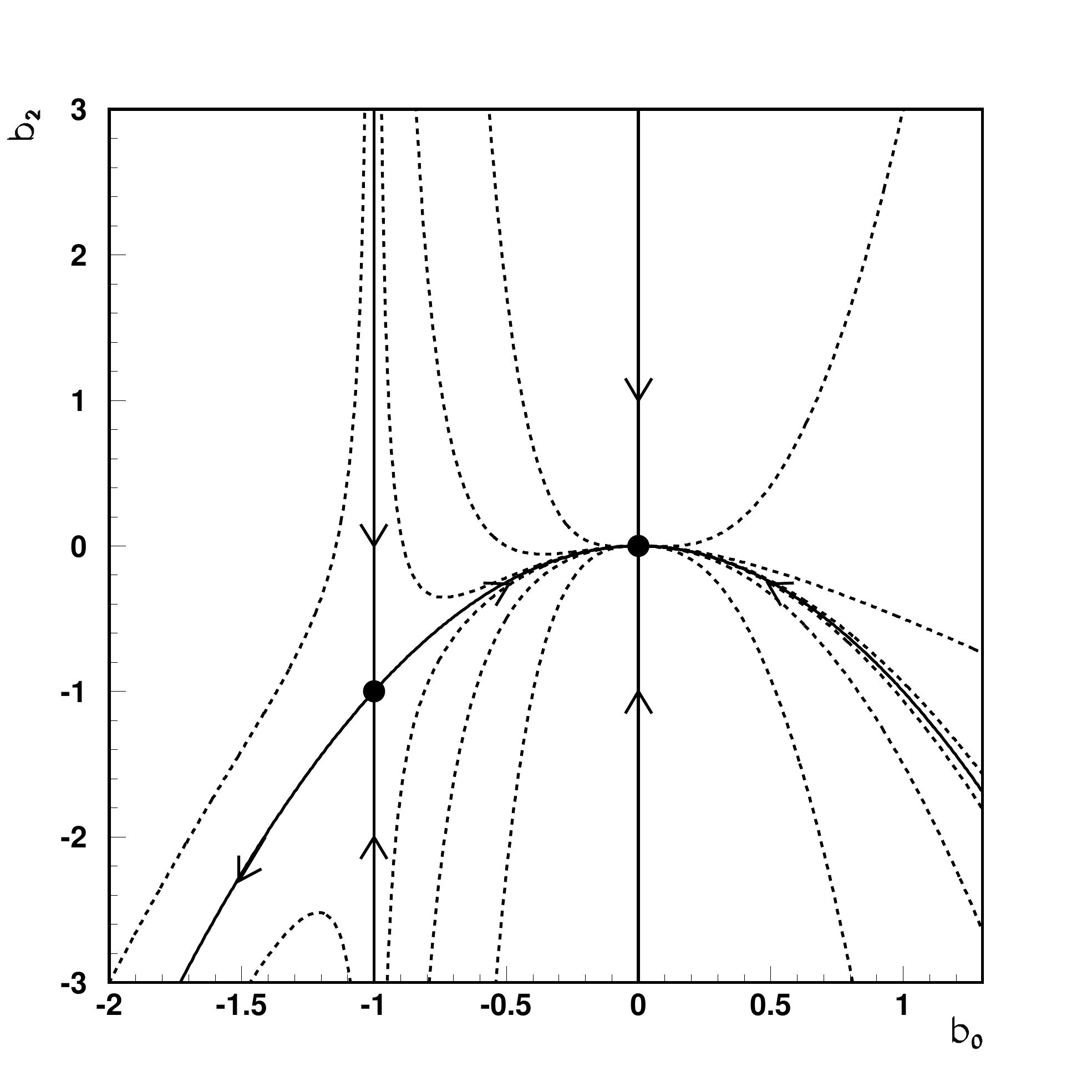}
\caption{RG flow of the potential $\hat V(\hat p,\Lambda)
=b_0(\Lambda)+b_2(\Lambda)\,\hat p^2+\cdots$.}
\end{figure}

Physical observables are given by the on-shell $K$-matrix. Returning to
physical units, this is
\begin{equation}
\frac{1}{K(p)}=\frac{M}{2\pi^2}\sum_{n=0}^\infty C_n\, p^{2n},
\end{equation}
where the $C_n$ are the coefficients of the RG eigenfunctions in $1/\hat V$.
Comparing this with
\begin{equation}
\frac{1}{K(p)}=-\,\frac{Mp}{4\pi}\left(-\,\frac{1}{a}
+\frac{1}{2}\,r_e\,p^2+\cdots\right),
\end{equation}
we see that this expansion is just the effective-range expansion 
(Bethe 1949; Blatt \& Jackson 1949). 
Note that the terms in the expansion of our effective theory correspond 
directly to scattering observables. 

The discussion above deals only with energy-dependent perturbations around the 
nontrivial fixed point. There are also RG eigenfunctions that depend on the
off-shell momenta. However, in contrast to the expansion around the trivial
point, these do not appear at the same orders as the corresponding on-shell
terms (Birse \textit{et al.}~1999). The momentum-dependent terms have larger 
scaling dimensions and so appear at higher orders in the EFT.

It is worth noting that the promotion of terms in the expansion
around the unitary limit can be understood from the form of the wave 
functions at short distances. Two particles in the unitary 
limit are described by irregular solutions of the Schr\"odinger equation. 
At small radii (in $S$ waves) these behave as $\psi(r)\propto r^{-1}$.
Any cutoff smears a contact interaction over range $R\sim\Lambda^{-1}$.
If we require observables to be independent of $\Lambda$, we therefore need 
the extra factor of $\Lambda^{-2}$ in the interaction to cancel the 
cutoff dependence from $|\psi(R)|^2\propto \Lambda^2$ in its matrix elements.
This provides the ``anomalous dimension" of $-2$ for the on-shell contact
interactions.

\section{Long-range forces}

Exactly the same techniques can be applied to renormalise short-range 
interactions in systems with known long-range forces.
Details can be found in the papers of Barford \& Birse (2003, 2005) and 
Birse (2006a, 2006b). The parameters of the resulting EFT again have 
a direct connection to scattering observables, either via a distorted-wave 
Born expansion (for weakly interacting systems) or via a distorted-wave 
effective-range expansion (for strong short-range interactions)
(Bethe, 1949; van Haeringen \& Kok 1982; Badalyan \textit{et al.}~1982).

The simplest example of these is the $1/r^2$ centrifugal barrier in partial waves
with nonzero angular momentum $L$ (Barford \& Birse, 2003). This leads to wave 
functions that behave as $r^L$ for small $r$ and so, near the trivial fixed 
point, short-range interactions scale with an additional power of $\Lambda^{2L}$ 
compared to $S$ waves. This matches exactly with the number of derivatives
needed to produce a contact interaction with a nonzero contribution in these 
waves.

The RG flow in these higher partial waves can also end up at fixed points with 
multiple unstable directions. However these turn out not to be physically
realisable, even with enough fine tuning: either the wave functions are not 
normalisable (Kaplan \textit{et al.}, 2009) or causality is not respected
(Hammer \& Lee, 2009, 2010).

In three-body systems with short-range pairwise interactions, long-range
forces are generated by exchange of one particle between an interacting
pair and the third particle. Close to the unitary limit, these can lead to
scaling behaviour of three-body interactions that is quite different  
from what naive dimensional analysis would suggest. The reasons for this
can be seen most clearly in position space if we work in hyperspherical
coordinates. At short distances the `particle-exchange" potential has
a $1/R^2$ form, where $R$ is the hyperradius (the radial coordinate that
is zero when all three particles coincide).

In three-body channels where this long-range potential is repulsive, it 
acts just like a centrifugal term but with a noninteger ``angular momentum".
This leads to wave functions that vanish as powers of the hyperradius as 
it tends to zero and hence to irrational anomalous dimensions
for the three-body contact interactions (Birse, 2006b). The values for
these have also been obtained from a momentum-space treatment (Griesshammer, 
2005) and by Werner and Castin (2006a, 2006b) from the eigenvalues of 
three-body systems trapped in a harmonic oscillator potential.

Systems of three bosons or three distinct fermions with a completely symmetric
spatial wave function behave very differently. Close to the unitary limit,
these display the towers of geometrically spaced bound states known as the 
Efimov effect (Efimov, 1971, 1979). This is a consequence of the attractive
$1/R^2$ potential in these systems, which leads to wave functions with the form
$\psi(R)\propto R^{-2\pm {\rm i}s_0}$, with $s_0\simeq 1.006$ as $R\rightarrow 0$.
As a result, the leading three-body force is promoted to a marginal term, of 
order $Q^{-1}$. The oscillatory behaviour associated with the imaginary part of 
the exponent is the origin of the Efimov effect. It causes the RG flow in these
systems to tend to a limit cycle instead of a fixed point (Bedaque, Hammer \& van 
Kolck, 1999a, 1999b, 2000; G\l azek and Wilson, 2004; Barford \& Birse, 2005).

In the presence of the Coulomb potential, the scaling behaviour of the 
short-range interactions is the same as that described in the previous section 
for systems without long-range forces (Barford \& Birse, 2003, Ando \& Birse,
2008). This is because the $1/r$ singularity of the long-range potential
is not strong enough to change the power-law behaviour of the wave 
functions at short distances. For strongly interacting systems, such as $pp$ 
scattering, the resulting EFT embodies the Coulomb distorted-wave 
effective-range expansion (Kong \& Ravndal, 1999, 2000).

Perhaps the most long-range important potential for nuclear physics is one-pion 
exchange (OPE). Although formally of order $Q^0$ within the framework of ChPT,
the large value of the pion-nucleon coupling means that it plays a central
role in nuclear forces. The unnaturally small value of the scale 
$\lambda_{\pi{\scriptscriptstyle NN}}$ associated with OPE suggests that it
should be added to our list of low-energy scales. This implies that OPE
should be treated nonperturbatively.

The central piece of OPE is the only one that contributes to scattering 
in spin-singlet waves. Like the Coulomb potential, this has a $1/r$ singularity, 
and so it does not alter the power-law forms of the wave functions at small $r$. 
The scattering in singlet waves with $L\geq 1$ is weak, and so the 
corresponding effective potential can be expanded using naive dimensional
analysis. In contrast, the $^1S_0$ channel has a low-energy virtual state
and so the expansion of its short-range potential is like the one around the 
unitary fixed point.

The tensor piece of OPE is important in spin-triplet waves. It has a 
much stronger, $1/r^3$, singularity at the origin. The resulting 
short-distance wave functions have the form $\psi(r)\propto r^{-1/4}$, 
multiplied by either a sine or an exponential of 
$(\lambda_{\pi{\scriptscriptstyle NN}}r)^{-1/2}$. As a result, short-range 
interactions are strongly promoted in these waves and so a new power 
counting needed, as observed by Nogga \textit{et al.} (2005). 
An RG analysis shows that the leading contact interaction is of order 
$Q^{-1/2}$ in waves with $L=1$ or 2 (Birse, 2006a). In the $^3S_1$ wave, 
there is a further enhancement of the short-range interactions, analogous 
to that in the $^1S_0$ wave, associated with the deuteron bound state.

Turning now to three-nucleon forces, two-pion exchange interactions are  
purely long-range and so are not renormalised by the effects discussed
here. In contrast, other interactions, involving two- or three-nucleon 
contact operators, are affected by the short-distance behaviour of the 
wave functions. For example, one-pion exchange terms of the type 
represented by figure~5, contain two-nucleon-one-pion contact operators
and these are substantially enhanced if either the initial or final
pair is in an $S$ wave.
\begin{figure}[h]
\begin{center}
\includegraphics[width=3cm]{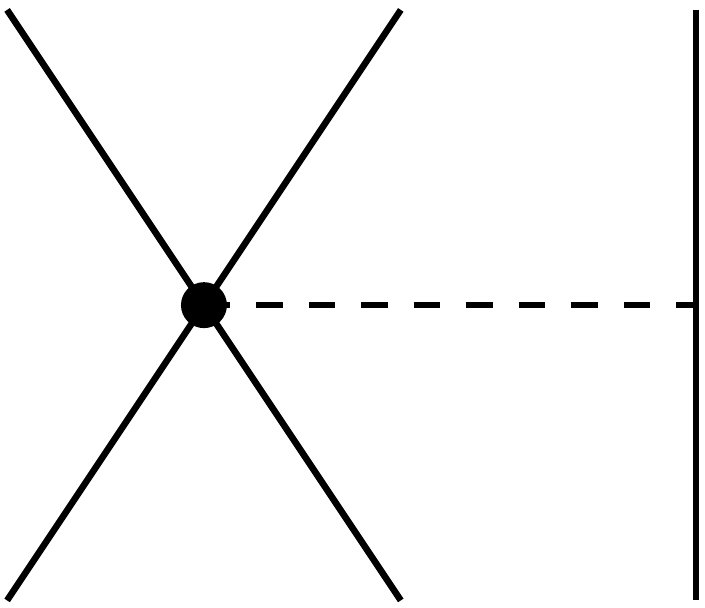}
\end{center}
\caption{A three-body OPE interaction.}
\end{figure}
The most important of these operators is one that couples the $^1S_0$ and 
$^3S_1$ NN channels This is promoted to order $Q^{5/4}$ by the 
nonperturbative treatment of the two-body forces in these channels
In addition, there can be strong promotion of operators that couple $S$
and $P$ waves, and of those that couple various combinations of $P$ and
$D$ waves. The effect of tensor OPE on three-body contact interactions
is currently unknown. The leading term is expected to be promoted, albeit 
less dramatically than in the case of pure short-range forces. Determining the 
anomalous dimension for this force will entail solving the three-body problem 
with $1/r^3$ two-body potentials. I have recently summarised the results 
obtained from of a range of RG studies of two- and three-nucleon 
forces (Birse, 2009).

Finally I should mention four-body systems. So far at least, these have not yielded 
to a detailed RG analysis of the sort outlined here. Numerical treatments of these 
systems have found no signs of promotion of four-body forces to relevant or marginal 
terms (Platter, Hammer \& Meissner, 2004, 2005; Platter \& Hammer, 2007) but the 
anomalous dimensions of these forces remain to be determined. In this context, the 
RG for the Legendre-transfor\-med effective action (Wetterich 1993) may provide 
some help. Applications of this rely on truncations of the effective action to a 
small number of local terms and, while less rigorous than truncations based on a 
specific power counting, these have proved remarkably successful for a wide variety 
of systems (Berges \textit{et al.}~2002) . Recently, the first applications have 
been made to few-body systems that make it possible to estimate the scaling dimensions 
of three- and four-body forces in both bosonic and fermionic systems close to the 
unitary limit (Schmidt \& Moroz, 2010; Birse \textit{et al.}, 2010b).

\section*{Acknowledgments}

I am grateful to my collaborators, especially T. Barford, J. McGovern, D. Phillips, 
and K. Richardson, for their contributions to the ideas summarised here. I must also
give special thanks to Y. Meurice and S.-W. Tsai, not only for the organisation
of the 2010 INT Workshop on ``New applications of the renormalization group
method" but also for their repeated encouragements to keep working on this 
review. I am indebted to the Institute for Nuclear Theory for its hospitality 
and support during several visits over the last two years.
This work was supported by the UK STFC under grant ST/F012047/1.

\end{document}